\documentstyle[aps,prd]{revtex} 

\draft

\begin{document}
\twocolumn[\hsize\textwidth\columnwidth\hsize\csname
@twocolumnfalse\endcsname 

\title{Electron-, Mu-, and Tau-Number Conservation 
in a Supernova Core}

\author{Steen Hannestad}
\address{NORDITA, Blegdamsvej 17, 2100 Copenhagen, Denmark}

\author{H.-Thomas Janka}
\address{Max-Planck-Institut f\"ur Astrophysik, 
Karl-Schwarzschild-Str.~1, 85740 Garching, Germany}

\author{Georg~G.~Raffelt} 
\address{Max-Planck-Institut f\"ur Physik 
(Werner-Heisenberg-Institut), 
F\"ohringer Ring 6, 80805 M\"unchen, Germany\\
and TECHNION, Israel Institute of Technology,
Haifa 32000, Israel}

\author{G\"unter Sigl} 
\address{D\'epartement d'Astrophysique Relativiste et de Cosmologie,\\ 
UMR 8629 du CNRS, Observatoire de Paris, 5 place J.~Janssen, 
92195 Meudon Cedex, France}

\date{8 December 1999, revised 19 May 2000}

\maketitle
                    
\begin{abstract}
  We study if the neutrino mixing parameters suggested by the
  atmospheric neutrino anomaly imply chemical equilibrium between
  $\mu$- and $\tau$-flavored leptons in a supernova (SN) core.  The
  initial flavor-conversion rate would indeed be fast if the
  $\nu_\mu$-$\nu_\tau$-mixing angle were not suppressed by {\it
  second-order\/} refractive effects. The neutrino diffusion
  coefficients are different for $\nu_\mu$, $\bar\nu_\mu$, $\nu_\tau$
  and $\bar\nu_\tau$ so that neutrino transport will create a net
  $\mu$ and $\tau$ lepton number density. This will typically lead to
  a situation where the usual first-order refractive effects dominate,
  further suppressing the rate of flavor conversion.  Altogether,
  neutrino refraction has the nontrivial consequence of guaranteeing
  the separate conservation of $e$, $\mu$, and $\tau$ lepton number in
  a SN core on the infall and cooling time scales, even when neutrino
  mixing angles are large.
\end{abstract}

\vskip2.0pc]


\section{Introduction}

The evidence for neutrino flavor oscillations from
atmospheric~\cite{SuperKamiokande} and solar neutrinos~\cite{solar} as
well as the LSND experiment~\cite{LSND} is now so compelling that the
debate in neutrino physics has fundamentally changed.  The current
experimental results point to a few very specific regions in the
parameter space of mass differences and mixing angles.  Therefore, the
experimental task at hand is to verify the proposed solutions by
independent means such as long-baseline oscillation experiments.  From
the astrophysical perspective, it is no longer an academic exercise to
investigate systematically if neutrino mixings with these specific
parameters lead to observable effects other than explaining the solar
neutrino deficit.

One consequence of the apparent violation of $e$, $\mu$, and $\tau$
flavor lepton number is that the leptons of a perfectly thermalized
system are no longer characterized by six independent chemical
potentials, but only by one for the charged leptons and one for the
neutrinos, i.e.\ $\mu_e=\mu_\mu=\mu_\tau$ and
$\mu_{\nu_e}=\mu_{\nu_\mu}=\mu_{\nu_\tau}$.  In practice one knows of
only two types of environment where neutrinos achieve thermal
equilibrium and where this effect could be important, the hot early
universe and the inner cores of collapsed stars.  In the standard
picture of the early universe, the neutrino chemical potentials are
very small so that flavor equilibration would not make much of a
difference except for the case of oscillations between active and
sterile neutrinos~\cite{Barbieri91}, a possibility that we will mostly
ignore.

In a supernova (SN) core, on the other hand, the electron lepton
number of the progenitor star's iron core is trapped, leading to
highly degenerate $e$ and $\nu_e$ distributions. If the trapped
electron lepton number were shared among all flavors, the impact on
the effective equation of state would be large.  Equilibration on the
infall time scale of a few hundred milliseconds could well be
disastrous for the SN explosion mechanism.  However, in a previous
paper~\cite{Raffelt93} two of us found that even for maximal neutrino
mixing, equilibrium between $\nu_e$ and the other flavors is not
achieved on the time scale of a few seconds unless $\delta m^2$
exceeds about $10^5~{\rm eV}^2$.  The rate of flavor conversion is
very slow because the mixing angle is strongly suppressed by neutrino
refractive effects. Since the current indications for neutrino
oscillations point to much smaller mass differences, the electron
lepton number in a SN core is almost perfectly conserved except for
diffusive or convective transfer to the stellar surface.

We presently study the corresponding conversion between $\nu_\mu$ and
$\nu_\tau$, which has not yet been investigated.  It is usually
assumed that the chemical potentials for these flavors vanish in a SN
core so that their distribution functions are equal, i.e.\ that flavor
equilibrium exists from the start.  However, the muon mass of
$106~{\rm MeV}$ is not very large, considering that typical
temperatures can exceed $T=30~{\rm MeV}$ and that the average thermal
energy of a relativistic particle is around $3T$. Beta equilibrium by
reactions of the form $\nu_\mu+n\leftrightarrow p+\mu^-$ implies the
condition $\Delta\mu\equiv\mu_n-\mu_p=\mu_\mu-\mu_{\nu_\mu}$, which is
familiar from the electron flavor.  The exact value of $\Delta \mu$
depends sensitively on the equation of state; a typical range is
50--100~MeV.  Since initially the trapped $\mu$ lepton number is zero,
and taking $\Delta\mu=50~{\rm MeV}$ and $T=30~{\rm MeV}$ as an
example, we find $\mu_\mu\approx32~{\rm MeV}$ and
$\mu_{\nu_\mu}\approx-18~{\rm MeV}$ so that there is a significant
excess of $\mu^-$ over $\mu^+$, compensated by an equal excess of
$\bar\nu_\mu$ over $\nu_\mu$.  On the other hand, the large value
$m_\tau=1777~{\rm MeV}$ completely suppresses the presence of $\tau$
leptons. Together with the absence of trapped $\tau$ lepton number
this implies $\mu_{\nu_\tau}=0$. The initial thermal distributions are
quite different between the $\mu$ and $\tau$ flavors!

Apart from these initial differences, it was recently recognized that
the opacities for $\nu_\mu$, $\bar\nu_\mu$, $\nu_\tau$ and
$\bar\nu_\tau$ are all different from each other if nucleon recoil
effects and muonic beta reactions are taken into account, leading to a
transient build-up of $\mu$ and $\tau$ lepton number as neutrinos
diffuse out of the star~\cite{Horowitz98}. Again, it is
important to understand if the $\mu$ and $\tau$ lepton numbers are
locally conserved on the diffusion time scale or if the condition
$\mu_{\nu_\mu}=\mu_{\nu_\tau}$ is enforced by flavor-conversion
processes.

The SuperKamiokande measurement of the atmospheric neutrino
anomaly~\cite{SuperKamiokande} suggests maximal
$\nu_\mu$-$\nu_\tau$-mixing and $\delta m^2$ roughly between $10^{-3}$
and $10^{-2}~{\rm eV}^2$. We now study if these mixing parameters lead
to chemical equilibrium between $\nu_\mu$ and $\nu_\tau$ on the
diffusion time scale of a few seconds.

\section{Flavor conversion}

Chemical equilibrium between charged leptons $\ell$ and neutrinos
$\nu_\ell$ of a given flavor is established by beta processes of the
form $\ell^-+p\leftrightarrow n+\nu_\ell$. They are ``infinitely
fast'' relative to all other time scales of interest.  Chemical
equilibrium between $\mu$- and $\tau$-flavored leptons, on the other
hand, is achieved by neutrino oscillations and by collisions which
break the coherence of mixed neutrino states.  The evolution of the
$\nu_\mu$-$\nu_\tau$-system is best described in terms of the
$2{\times}2$ density matrices $\rho_{\bf p}$ for each momentum ${\bf
  p}$.  In the flavor basis, the diagonal entries of $\rho_{\bf p}$
are the usual $\nu_\mu$ and $\nu_\tau$ occupation numbers,
respectively, while the off-diagonal entries represent phases produced
by oscillations. The evolution of $\rho_{\bf p}$, and of
${\bar\rho}_{\bf p}$ for anti-neutrinos, is governed by a generalized
Boltzmann collision equation which simultaneously includes the effects
of oscillations and collisions~\cite{RSS93,Sigl93}.

For our simple estimates, however, it will suffice to use a
``pedestrian'' method where the neutrino ensemble is represented by a
single mode with momentum ${\bf p}$ or energy $E\approx|{\bf p}|$.
Without collisions, the flavor content of
this mode would oscillate forever. Interactions with the medium,
however, destroy the coherence between the flavor components, leading
eventually to an equal, incoherent mixture.  Maximally mixed neutrinos
approach this state of flavor equilibrium with a
rate~\cite{Stodolsky87}
\begin{equation}\label{eq:gammaflavor}
\Gamma_{\rm flavor}
=\omega_{\rm vac}^2\, \frac{D}{D^2+(\delta V)^2},
\end{equation}
where
\begin{eqnarray}\label{eq:omegavac}
\omega_{\rm vac}\equiv\frac{\delta m^2}{2E}
&=&1.7\times10^{-11}~{\rm eV}\,\,\Delta_{3}\,E_{30}^{-1}\nonumber\\
&=&2.5\times10^{4}~{\rm s}^{-1}\,\,\Delta_{3}\,E_{30}^{-1}
\end{eqnarray}
is the oscillation frequency in vacuum.  Here, $\Delta_{3}\equiv\delta
m^2/10^{-3}~{\rm eV}^2$ represents the lower end of the mass range
implied by SuperKamiokande and $E_{30}\equiv E/30~{\rm MeV}$ where
30~MeV is a typical temperature in a SN core.  Further, $\delta V$ is
the energy difference between $\nu_\mu$ and $\nu_\tau$ of equal
momenta caused by the medium, i.e.\ $\delta V$ is the difference of
the medium's weak potential for our two neutrino flavors.

Finally, $D$ is the damping or decoherence rate, i.e.\ 
the rate by which interactions with the medium ``measure'' the flavor
content of a mixed neutrino state. Typically $D$ is of the order of the
neutrino collision rate, but the exact relationship between the two
quantities is not trivial.  For example, in a situation where one of
the neutrino flavors scatters with a rate $\Gamma_{\rm coll}$ while
the other is sterile, one finds $D=\Gamma_{\rm coll}/2$
\cite{Stodolsky87}.  Equation~(\ref{eq:gammaflavor}) applies in the
``strong damping limit'' defined by $D\gg\omega_{\rm vac}$, a
condition which is satisfied in our scenario.

Equation~(\ref{eq:gammaflavor}) is easily interpreted in two limiting
cases. For $\delta V=0$, the flavor content of a given state
oscillates as $\frac{1}{2}[1+\cos(\omega_{\rm vac} t)]$. The
oscillations are interrupted by those collisions which ``measure'' the
difference between $\nu_\mu$ and
$\nu_\tau$~\cite{RSS93,Sigl93,Stodolsky87}. If this ``measurement
rate'' or ``decoherence rate'' $D=\tau_D^{-1}$ is much larger than
$\omega_{\rm vac}$, the flavor oscillation is interrupted when
$\omega_{\rm vac}t\ll 1$ so that the flavor content evolves as
$(\omega_{\rm vac} t/2)^2$ until it is interrupted.  This happens with
a rate $\tau_D^{-1}$ so that the rate of flavor conversion must scale
as $(\omega_{\rm vac} \tau_D)^2 \tau_D^{-1}$ or
\begin{equation}
\Gamma_{\rm flavor}=\frac{\omega_{\rm vac}^2}{D}.
\end{equation}
In this case the flavor conversion rate decreases with increasing $D$,
and vanishes for infinite $D$.  This situation is known as the
``Quantum Zeno Paradox'' or ``Watched Pot Effect'' \cite{Stodolsky87}:
The neutrino remains ``frozen'' in its flavor state because it is
frequently ``watched'' or measured to be in this state by the
interactions with the medium.

A more familiar limiting case obtains when $|\delta V|\gg D$ so that
$D^2$ in the denominator of Eq.~(\ref{eq:gammaflavor}) can be
neglected.  Since $D\gg \omega_{\rm vac}$ by assumption, we also have
$|\delta V|\gg \omega_{\rm vac}$, implying that the oscillation
frequency is $\delta V$ instead of $\omega_{\rm vac}$ because the
energy difference between $\nu_\mu$ and $\nu_\tau$ of equal momenta is
now dominated by $\delta V$, not by $\omega_{\rm vac}$.  Since
$|\delta V|\gg D$, the collisions are rare relative to the oscillation
period.\footnote{Sometimes this situation is referred to as ``weak
  damping.'' However, we follow the convention of
  Ref.~\protect\cite{Stodolsky87} where ``weak damping'' means
  $D\ll\omega_{\rm vac}$.}  Averaging over an oscillation period, and
if we begin with one flavor, the average probability for the
appearance of the other is $\frac{1}{2}\sin^2(2\Theta)$ where $\Theta$
is the in-medium mixing angle. If the oscillations are interrupted
with a rate $\Gamma_{\rm coll}$, we have
\begin{equation}
\Gamma_{\rm flavor}= \sin^2(2\Theta)\,D
\end{equation}
if $D$ is interpreted as $\Gamma_{\rm coll}/2$.  For maximally mixed
neutrinos, the in-medium mixing angle is given by
\begin{equation}
\tan(2\Theta)=\frac{\omega_{\rm vac}}{\delta V}.
\end{equation}
Since $|\omega_{\rm vac}/\delta V|\ll1$ we have $\tan(2\Theta)
\approx\sin(2\Theta)$ so that we recover Eq.~(\ref{eq:gammaflavor}). A
large refractive energy difference between the flavors suppresses the
conversion rate, an effect which evidently can be interpreted as a
suppression of the mixing angle in the medium.

\section{Rate of Decoherence}

In order to estimate the rate of flavor conversion we are thus left
with the task of estimating the damping rate, $D$, and the refractive
effect $\delta V$ for the conditions of a SN core.  The largest
possible conversion rate obtains for $\delta V=0$ so that we first
consider this case.  If $\Gamma_{\rm flavor}$ were slow in this limit,
a further discussion of refractive effects would be unnecessary.

We thus begin by comparing $\Gamma_{\rm flavor}=
\omega_{\rm vac}^2 D^{-1}$ with the
diffusion rate $\tau_{\rm diffusion}^{-1}\approx1~{\rm s}^{-1}$,
\begin{equation}
\Gamma_{\rm flavor}\tau_{\rm diffusion}
=\frac{6.4\times10^8~{\rm s^{-1}}}{D}\,\frac{\Delta_3^2}{E_{30}^2},
\end{equation}
where we have used Eq.~(\ref{eq:omegavac}).  Evidently it is the
low-energy neutrino modes with their fast oscillation frequency which
are most effective for flavor conversion.

There are two conceptually different contributions to $D$.
Oscillations are interrupted by scattering processes which are
sensitive to the neutrino flavor such as $\nu_\mu+e^-\to\nu_e+\mu^-$,
a process which has no analogue for $\nu_\tau$ because of the large
$\tau$ mass.  In addition, a mixed neutrino can scatter to higher
energies, for example by $\nu+e^-\to e^-+\nu$, without interrupting
the oscillation process. However, after this ``up-scattering,'' the
probability for processes which do distinguish between the flavors
increases significantly since the cross sections for all neutrino
processes increase with energy. Moreover, at large energies the fast
beta process $\nu_\mu+n\to p+\mu^-$ becomes kinematically allowed.
Effectively, the up-scattering of a low-energy neutrino causes the
oscillations to be interrupted, even if the up-scattering process
itself is flavor-diagonal. Therefore, up to a numerical factor, the
decoherence rate $D$ of low-energy neutrinos is identical with their
scattering rate to higher energies.

The most important energy-changing process for neutrinos in a SN core
is the scattering on the nuclear medium.  Ignoring the subdominant
vector-current interaction, the differential cross-section for
neutral-current scattering of a neutrino of energy $E_1$ to energy
$E_2$ is
\begin{equation}\label{eq:differentialcrosssection}
\frac{d\sigma}{dE_2}=
\frac{3C_A^2\,G_F^2}{\pi}\, E_2^2\,\frac{S(E_1-E_2)}{2\pi}.
\end{equation}
Here, $S(\omega)$ is the dynamical structure function for the
axial-vector current interaction in the long-wavelength 
limit~\cite{Janka96}. In a
dilute medium $S(\omega)=2\pi \delta(\omega)$, leading to the usual
neutral-current elastic scattering cross section. In a nuclear medium,
nucleon-nucleon interactions cause $S(\omega)$ to be a broadly
smeared-out function which obeys the detailed-balancing condition
$S(-\omega)=S(\omega) e^{-\omega/T}$. Unless spin-spin correlations
and degeneracy effects are important, the structure function has the
norm $\int S(\omega)d\omega/2\pi=1$.

In contrast with elastic neutrino scattering processes, the cross
section Eq.~(\ref{eq:differentialcrosssection}) does not vanish for
small neutrino energies.  With $E_1=0$, the neutrino scattering rate
on nucleons is
\begin{equation}
\Gamma_{\rm nuc}(0)=
\frac{3C_A^2G_F^2}{\pi}\,T^2\,n_{B} 
\int_0^\infty dx\, x^2 \frac{T S(-Tx)}{2\pi} 
\end{equation}
where $n_{B}$ is the baryon (nucleon) number density.  With
$C_A=1.26/2$ for neutral-current processes, the coefficient before the
integral is
\begin{equation}
0.98\times10^8~{\rm s}^{-1}\,
\frac{\rho}{3\times10^{14}~{\rm g~cm^{-3}}}\,T_{30}^2, 
\end{equation}
where $T_{\rm 30}\equiv T/30~{\rm MeV}$.

The dimensionless integral vanishes if $S(\omega)$ is very narrow,
corresponding to the usual case of a vanishing elastic neutrino
scattering cross section at $E_1=0$.  The integral also vanishes when
it is very broad because $S(\omega)$ decreases at least exponentially
for large negative $\omega$. If $S(\omega)$ is normalized, and if it
is a smoothly varying broad function, the maximum possible value for
the integral expression is about $0.25$ which obtains when the width
of $S(\omega)$ is of order the temperature, probably corresponding to
realistic SN conditions.  The rate gets reduced by nucleon degeneracy
effects and spin-spin anticorrelations. Therefore, for low-energy
neutrinos the upscattering rate is probably not larger than a few
times $10^7~{\rm s}^{-1}$.  For low-energy neutrinos, this rate is
larger than any other up-scattering process that we could identify
such as neutrino-electron or neutrino-neutrino scattering.

Below the muon production threshold in beta processes, the most
important reactions which distinguish between $\nu_\mu$ and $\nu_\tau$
appear to be $\nu_\mu + e^- \to \mu^- + \nu_e$ and
$\bar\nu_\mu+\mu^-\to e^-+\bar\nu_e$. We find that for typical
conditions in a SN core and for neutrino energies around $T$ the rate
is at most a few times $10^7~{\rm s}^{-1}$ and thus comparable to
neutrino nucleon scattering.

Comparing these numbers with Eq.~(\ref{eq:omegavac}) reveals that
$D\gg\omega_{\rm vac}$. Therefore, the strong damping limit indeed
applies as had been assumed earlier.

The actual decoherence rate $D$ is smaller than our estimates of
the corresponding collision rates. We have already mentioned
that $D$ is {\it half\/} the collision rate
in a situation where one flavor scatters while the other is
sterile~\cite{Stodolsky87}.
Altogether, we believe that
\begin{equation}
D\alt3\times10^{7}~{\rm s}^{-1}
\end{equation}
is a realistic upper limit for typical SN conditions of
$\rho=3\times10^{14}~{\rm g~cm^{-3}}$ and $T=30~{\rm MeV}$, implying
\begin{equation}
\Gamma_{\rm flavor}\tau_{\rm diffusion}
\agt 20\,\Delta_{3}^2\,E_{30}^{-2}.
\end{equation}
Averaging this expression over a thermal neutrino distribution,
we note that $\langle E^{-2}\rangle=0.38\,T^{-2}$ so that
we finally conclude that for typical SN conditions
\begin{equation}
\langle \Gamma_{\rm flavor}\rangle\tau_{\rm diffusion}
\agt 10\,\Delta_{3}^2.
\end{equation}
Near the upper end of the mass range suggested by SuperKamiokande we
have $\delta m^2=10^{-2}~{\rm eV}^2$ or $\Delta_{3}=10$, implying that
the flavor conversion rate increases by two orders of magnitude
relative to our estimate.

In summary, it appears that the flavor conversion between $\nu_\mu$
and $\nu_\tau$ in a SN core would be much faster than the diffusion
time scale of about one second if refractive effects could be ignored.

\section{Refractive Effects}

Turning to neutrino refraction in the SN medium, we begin with the
usual lowest-order weak potential,
\begin{equation}
\delta V^{(1)}=\sqrt{2}\, G_F \left(\Delta n_{\mu}+\Delta n_{\nu_\mu}
-\Delta n_{\tau}-\Delta n_{\nu_\tau}\right),
\end{equation}
where $G_F$ is the Fermi constant while $\Delta n_j$ stands for the
number density of particle $j$ minus the density of anti-particles.
Since the total $\mu$ and $\tau$ lepton number trapped in a SN core is
zero, we have initially $\delta V^{(1)}=0$. Therefore, the flavor
conversion rate would seem to be initially fast until a significant
weak potential difference has built up.

However, second-order effects can not be neglected. At one-loop level,
the neutral-current interactions of $\nu_\mu$ and $\nu_\tau$ with the
nucleons are not identical because of the different charged-lepton
masses in the loop.  The second-order energy difference in a normal
medium was found to be~\cite{Botella}
\begin{eqnarray}\label{eq:secondorder}
|\delta V^{(2)}|&=&\frac{3 G_F^2 m_\tau^2}{2\pi^2}\,n_B\,
\left[\ln\left(\frac{m_W^2}{m_\tau^2}\right)-1+\frac{Y_n}{3}\right]
\nonumber\\
&=&6.3\times10^{-4}~{\rm eV}\,
\frac{\rho}{3{\times}10^{14}~{\rm g/cm^{3}}}\,
\frac{6.61+Y_n/3}{7}\nonumber\\
\end{eqnarray}
where $n_B$ is the baryon density and $Y_n$ the neutron number
fraction. With $6.3\times10^{-4}~{\rm eV}=9.6\times10^{11}~{\rm
  s}^{-1}$ we find $|\delta V^{(2)}|\gg D\gg \omega_{\rm vac}$ and
that even initially the flavor conversion time scale (the inverse of
$\Gamma_{\rm flavor}$) far exceeds the one for diffusion.
This must be the only example where radiative corrections to the
neutrino refractive index are of direct practical relevance! 

We stress, however, that $\delta V^{(2)}$ is by no means exotically
large. Its importance in the present context derives from the
cancellation of $\delta V^{(1)}$ and from the smallness of
$\omega_{\rm vac}$. For example, the in-medium mixing angle in the
present case is
\begin{equation}
\tan(2\Theta)=2.6\times10^{-8}\,\,
\frac{3{\times}10^{14}~{\rm g/cm^{3}}}{\rho}\,
\frac{7}{6.61+Y_n/3},
\end{equation}
i.e.\ second-order refractive effects suppress the mixing angle by
about eight orders of magnitude!  The real surprise is that
$\omega_{\rm vac}$, despite its smallness, is large enough to cause
flavor equilibrium if it were not for the refractive suppression of
the mixing angle.

Another second-order contribution to $\delta V$ arises from the
low-energy tail of the $W^\pm$ and $Z^0$ resonance in neutrino forward
scattering on other neutrinos and charged leptons~\cite{Notzold88}.
This term is the dominant second-order correction in the early
universe where the particle-antiparticle asymmetries are small, but in
a SN core Eq.~(\ref{eq:secondorder}) is more important because of the
huge baryon density.

Once diffusion processes begin to build up a net $\mu$ or $\tau$
lepton number in the SN core, the first-order refractive effect
becomes important. Its magnitude is understood if we calculate the
refractive energy shift caused by neutrinos with a chemical potential
$\mu_\nu\ll T$ which we express as $\eta_\nu=\mu_\nu/T$. To lowest
order we have $\Delta n_\nu=T^3\eta_\nu/6+{\cal O}(\eta_\nu^2)$ so
that
\begin{equation}\label{eq:neutrinoshift}
\delta V^{(1)}_\nu=0.074~{\rm eV}\,\,T_{30}^3\eta_\nu.
\end{equation}
Likewise, for muons we find $\Delta
n_{\mu}=(T^3/\pi^2)\,f(m_\mu/T)\,\eta_\mu$ where the function
$f(m_\mu/T)$ is 1.077 for $m_\mu/T=3$. Therefore, muons provide a
similar energy shift. Altogether, if we use $\eta$ as some
characteristic chemical potential for the muons and neutrinos,
the first-order energy shift is given by 
Eq.~(\ref{eq:neutrinoshift}), leading to an in-medium mixing
angle of
\begin{equation}\label{eq:mixingangle}
\tan(2\Theta)=2.2\times10^{-10}\,\,
\frac{\Delta_3}{\eta\,E_{30} T_{30}^3}.
\end{equation}
Again, the mixing angle is hugely suppressed unless $\eta$ is finely
tuned to zero.  Considering that $\Gamma_{\rm flavor}=\sin^2(2\Theta)
D$ the suppression of the conversion rate is quadratic in the small
in-medium mixing angle. For $D={\cal O}(10^{7}~{\rm s}^{-1})$ even a
tiny value for $\eta$ is enough to suppress flavor conversion
completely.

As $\mu$ and $\tau$ lepton number builds up it may happen that the first
and second-order contributions to $\delta V$ cancel for suitable
conditions, allowing briefly for a fast flavor conversion rate.
However, the slightest deviation from this condition again suppresses
the mixing angle and thus quenches any further flavor conversion.  Any
attempt to reach flavor equilibrium by neutrino oscillations is
robustly self-quenching.

We have mostly studied typical core conditions in the SN.  In regions
below the neutrino sphere the density is about three orders of
magnitude less than our benchmark value of nuclear density, and the
temperature may be a factor of 5 smaller than our standard figure of
30~MeV. For such conditions, the presence of muons is strongly
suppressed so that the decoherence rate $D$ is much smaller than what
we have estimated for average core conditions.  On the other hand, the
in-medium mixing-angle Eq.~(\ref{eq:mixingangle}) is still very small
if we use $T_{30}=0.2$ and $E_{30}=0.2$ as characteristic values near
the neutrino sphere. Therefore, our conclusion that flavor equilibrium
cannot be achieved applies to conditions throughout the SN core.

\section{Conclusions}

The chemical equilibration between the $\mu$ and $\tau$ flavors in a
SN core by neutrino oscillations would be fast on the neutrino
diffusion time scale if the refractive energy shift $\delta V$ between
$\nu_\mu$ and $\nu_\tau$ were small.  The usual first-order
contribution indeed vanishes due to the lack of trapped $\mu$- and
$\tau$-lepton number, but second-order contributions are large enough
to suppress flavor conversion.

In the course of neutrino transport to the stellar surface, $\mu$- and
$\tau$-lepton number will build up due to the differences between the
$\nu_\mu$, $\bar\nu_\mu$, $\nu_\tau$, $\bar\nu_\tau$ diffusion
constants~\cite{Horowitz98}. It is conceivable that
local conditions can be reached where the refractive energy shift
$\delta V$ vanishes to all orders.  While this cancellation would
momentarily lead to a fast rate of flavor conversion, the resulting
re-distribution of $\mu$ and $\tau$ lepton number quickly produces a
$\delta V$ so large that the conversion rate is suppressed again. The
equilibrium condition $\mu_{\nu_\mu}=\mu_{\nu_\tau}$ implies a $\delta
V$ so large that it can never be reached, i.e.\ the flavor conversion
process is inevitably self-quenching.

For the small neutrino mass differences indicated by the experiments,
the lack of conversion between $\nu_e$ and the other flavors is much
easier to understand because the large amount of trapped
electron-lepton number causes $\delta V^{(1)}$ to be so large that the
in-medium mixing angle is easily seen to be vastly
suppressed~\cite{Raffelt93}.

The phenomenon of neutrino mixing and neutrino oscillations may have a
variety of astrophysical consequences which need to be explored.  In
the past, most of the attention was focussed on the possibility of
detecting evidence for neutrino oscillations in the astrophysical
context.  However, as neutrino oscillations become more and more
experimentally established, the problem of unravelling the neutrino
mass matrix may become a less pressing astrophysical preoccupation
than the reverse question: given the experimentally measured neutrino
parameters, is the effect of flavor violation important or ignorable
in a given environment?

We think it is intriguing that the core of a SN is protected from the
consequences of flavor-lepton number violation by the phenomenon of
neutrino refraction and that second-order effects have to be included.
While the significance of the weak-interaction potential for the
resonant enhancement of oscillations in the spirit of the MSW effect
has been widely acknowledged, the suppression of oscillations in a SN
core is another consequence of neutrino refraction with real and
important astrophysical ramifications.

 
\section*{Acknowledgments}

We thank Leo Stodolsky for reading the manuscript and several helpful
comments.  In Munich, this work was partly supported by the Deut\-sche
For\-schungs\-ge\-mein\-schaft under grant No.\ SFB 375. In 
Copenhagen, it was
supported by a grant from the Carlsberg Foundation.


\end{document}